\documentstyle[12pt]{article}

\newcommand{\al}{\alpha}
\newcommand{\be}{\begin{equation}}
\newcommand{\ee}{\end{equation}}
\newcommand{\prt}{\partial}
\newcommand{\lbd}{\lambda}
\newcommand{\om}{\omega}
\newcommand{\gm}{\gamma}
\newcommand{\bt}{\beta}

\begin{document}

\begin{center}
{\Large{\bf Magnetic Semiconfinement of Neutral Atoms} \\ [5mm]
V.I. Yukalov$^1$ and E.P. Yukalova$^2$} \\ [3mm]

{\it $^1$Bogolubov Laboratory of Theoretical Physics \\
Joint Institute for Nuclear Research, Dubna 141980, Russia \\
and \\
$^2$Laboratory of Informational Technologies \\
Joint Institute for Nuclear Research, Dubna 141980, Russia}

\end{center}

\vskip 2cm

\begin{abstract}

A mechanism for creating well-collimated beams of neutral particles or 
atoms with spins is studied. The consideration is accomplished for a 
general realistic case, taking into account: (i) the finiteness of a 
cylindrical trap where the atoms to be shot out, are stored; (ii) the
possibility of manipulating the trap magnetic fields in order to form
different anisotropic field configurations; (iii) the presence of
gravity curving atomic trajectories.

\end{abstract}

\vskip 1cm


\vskip 1cm


\newpage

\section{Introduction}

Charged particles and ions are known to be easily manipulated by means 
of electromagnetic fields. The situation is more complicated for neutral
particles and atoms. In what follows, we shall often employ, for short, 
the term "atom" for both elementary neutral particles as well as for 
composite systems, such as atoms or molecules. Neutral
atoms can be successfully confined in
different traps (see reviews [1--3]). Recently [4], neutrons were confined
in a magnetic trap during the time of about 800 seconds, that is, around 
13 minutes; this lifetime being mainly limited by the neutron beta decay
into an electron, proton, and anti-neutrino, rather than by the trap 
losses. The formation of directed beams of neutral atoms is usually done 
in one of the two following ways. The first way can be called 
{\it mechanical}, since it requires the usage of mechanical collimators,
selecting atoms from an isotropic distribution, after which the selected 
particles are released through tubes with a high pressure difference
between the ends, as is done in molecular beam masers [5,6]. Another 
mechanism may be named {\it resonant}, because it is based on the action
of laser beams on resonant atomic transitions [1--3].

A novel mechanism for creating directed beams of neutral atoms has been
recently advanced [7,8]. This mechanism can be termed {\it magnetic},
since it involves magnetic fields, together with a special
initial polarization of atomic spins. A short qualitative description of 
the suggested {\it magnetic semiconfinement of atoms} is directly connected
with its name: This is an effect, when neutral atoms, stored in a trap, begin
moving  in one direction, forming a well-collimated beam, at the same time 
being confined from the opposite side.

The ability to form directed well-collimated beams of neutral atoms is
very important for various applications. For example, this is necessary 
for the functioning of atom lasers [2,3]. This can also be used for 
studying the coherence of trapped Bose-Einstein condensates [3,9], for
considering the relative motion of binary mixtures [10], for measuring
the one-particle density matrix by the scattering of fast atoms [11], 
and so on.

In order that a method of creating atomic beams be practical, it is 
necessary to check its feasibility under a realistic situation. And this 
is the aim of the present paper to generalize the consideration of the
magnetic mechanism for creating atomic beams [7,8] by including the
following really existing essential factors.

\vskip 2mm

(i) {\it Trap shape}. Atoms to be shot out are, first, stored in a trap
having a finite size and a particular shape [1--3]. The formation of an
atomic beam is accomplished inside the trap whose shape, therefore, is 
of importance.

\vskip 2mm

(ii) {\it Field anisotropy}. Fields governing the dynamics of atoms can 
be prepared in different configurations [1--3,12]. It is useful to include 
in the consideration additional parameters regulating the anisotropy of 
field configurations, which would help in choosing optimal conditions 
for the formation of beams.

\vskip 2mm

(iii) {\it Influence of gravity}. Atoms, contrary to photons, possess 
mass. Hence atoms are subject to gravity that may strongly influence 
their dynamics [3,13]. Therefore one has to accurately take into account
the presence of gravity.

\section{Semiconfining regime of motion}

Let us consider a system of atoms, each having mass $m_0$, magnetic moment
$\mu_0$, and spin $S$. Introduce the quantum-mechanical averages for the
spatial position of an atom ${\bf r}=\{ r^\al\}$, with $\al=x,y,z$, and
for a spin ${\bf S}=\{ S^\al\}$. Denote the gravitational acceleration
as ${\bf g}=\{ g^\al\}$ and the force due to interatomic interactions
by ${\bf f}=\{ f^\al\}$. The evolution equations for the averages ${\bf r}$
and ${\bf S}$, in the semiclassical approach [14,15], can be presented in 
the form
\be
\frac{d^2 r^\al}{d t^2} = \frac{\mu_0}{m_0}\; {\bf S} \cdot
\frac{\prt{\bf B}}{\prt r^\al} + g^\al +
\frac{f^\al}{m_0} \; ,
\ee
where ${\bf B}$ is a magnetic field acting on an atom, and
\be
\frac{d{\bf S}}{dt}  =\frac{\mu_0}{\hbar}\; {\bf S}\times {\bf B} \; .
\ee
The total magnetic field of a trap can be written as a sum
\be
{\bf B} = {\bf B}_1 + {\bf B}_2
\ee
of a time-invariant trapping field ${\bf B}_1$ and an alternating field 
${\bf B}_2$. As a trapping field, one often uses the quadrupole field
\be
{\bf B}_1 = B_1'\left ( x {\bf e}_x + y {\bf e}_y + 
\lbd z {\bf e}_z  \right ) \; ,
\ee
typical of magnetic traps [1--3], where ${\bf e}_\al$ are unitary vectors,
$B_1'>0$ is a field gradient and $\lbd$ is an anisotropy parameter that
can be varied. The alternating field
\be
{\bf B}_2 =  B_2 \left ( h_x {\bf e}_x + h_y {\bf e}_y \right ) \; ,
\ee
with $h_\al=h_\al(t)$ and $h_x^2 + h_y^2 = 1$, is a transerse field serving 
for the stabilization of atomic confinement.

As is clear from the formulas (4) and (5), these magnetic fields describe a 
generalized TOP (time orbiting potential) trap. The fields specific for the 
latter follow if in Eq. (4) one sets $\lbd=-2$ and in Eq. (5) a rotating 
transverse field is assumed. We accept here a more general form of the trap 
fields in order to stress that the considered mechanism of beam creation 
does not compulsory require the usage of the TOP traps but can be realized 
in other traps as well. Moreover, contrary to the TOP trap, with a fixed 
anisotropy, there exists a rich variety of other traps, where the field 
anisotropy can be varied in a wide diapason, with the ratio of radial to 
axial frequencies reaching a factor of 100. More details on possible field 
configurations can be found in reviews [1--3]. Modelling here the trap 
anisotropy by the parameter $\lbd$, we demonstrate the main idea that the 
semiconfining regime of motion can always be optimized by adjusting the 
corresponding configuration of trap fields.

In what follows, it is convenient to pass to the dimensionless space variable
${\bf r}=\{ x,y,z\}$ with the dimensionless components
$$
x \equiv \frac{r^x}{R_0} \; , \qquad y \equiv \frac{r^y}{R_0} \; ,
\qquad z \equiv \frac{r^z}{R_0} 
$$
measured in units of the characteristic length
\be
R_0 \equiv \frac{B_2}{B_1'} \; .
\ee
The latter is called the death circle, which describes the position of the 
oscillating zero magnetic field. The length (6) limits the size of the 
trapable cloud whose actual radius is slightly smaller [1--3]. Introduce 
the characteristic frequency of atomic motion, $\om_1$, and that of 
spin motion, $\om_2$, by the equalities
\be
\om_1^2 \equiv \frac{\mu_0 B_1'}{m_0 R_0} \; , \qquad
\om_2 \equiv \frac{\mu_0 B_2}{\hbar} \; .
\ee
And let us employ the notation
\be
{\bf G} \equiv \frac{{\bf g}}{R_0 \om_1^2} \; , \qquad
\gm{\vec \xi} \equiv \frac{{\bf f}}{m_0 R_0} \; ,
\ee
where ${\bf G}$ means the dimensionless gravitational force and $\gm$ implies
a collision rate. Atomic collisions can be modelled by a random variable
${\vec\xi}=\{\xi_\al(t)\}$ defined by means of the stochastic averages
\be
\ll\xi_\al(t)\gg \; = 0 \; , \qquad \ll \xi_\al(t)\;\xi_\bt(t')\gg \; =
2 D_\al\delta_{\al\bt}\; \delta(t-t') \; ,
\ee
in which $D_\al$ is a diffusion rate.

With this notation, the evolution equation (1) takes the form of the stochastic 
differential equation
\be
\frac{d^2{\bf r}}{d t^2} = \om_1^2 \left ( S^x {\bf e}_x + S^y {\bf e}_y +
\lbd S^z {\bf e}_z +{\bf G}\right ) +\gm{\bf\xi} \; ,
\ee
and the spin equation (2) can be transformed as
\be
\frac{d{\bf S}}{dt} = \om_2 \hat A{\bf S} \; ,
\ee
where $\hat A =[A_{\al\bt}]$ is an antisymmetric matrix with the elements
$$
A_{\al\al} = 0 \; , \qquad A_{\al\bt} = - A_{\bt\al} \; ,
$$
$$
A_{12} =\lbd z \; , \qquad A_{23} = x +h_x \; , \qquad A_{31}= y + h_y \; .
$$

The evolution equations can be simplified by taking account of the existence 
of small parameters
\be
\left | \frac{\om_1}{\om_2}\right | \ll 1 \; , \qquad
\left | \frac{\om}{\om_2}\right | \ll 1 \; , \qquad
\left | \frac{\gm}{\om_2}\right | \ll 1 \; , 
\ee
where the notation
\be
\om \equiv \max_t \left | \frac{d}{dt}\; {\bf h}(t)\right |
\ee
is used. These inequalities are usually valid for realistic traps [1--3]. Then 
the evolution equations can be reduced to the consideration of atomic motion 
on the center manifold by employing the scale separation approach [13,16], 
which involves a generalization of averaging techniques [17] to stochastic 
equations.

Under conditions (12), the matrix $\hat A$ can be treated as a quasi-invariant
[13,18] with respect to the fastly oscillating spin ${\bf S}$. Then the 
solution to Eq. (11) can be presented as a sum
\be
{\bf S}(t) = \sum_{j=1}^3 C_i{\bf b}_j(t) \exp\{ \om_2 a_j(t) t\}
\ee
over the eigenvectors ${\bf b}_j={\bf b}_j(t)$ of the matrix $\hat A$, where
$a_j=a_j(t)$ are the related eigenvalues, and the coefficients $C_j$ are given 
by the initial conditions, so that
$$
C_j ={\bf S}(0)\cdot {\bf b}_j(0) \; ,
$$
$$
{\bf b}_j =
\frac{(A_{12}A_{23} - a_j A_{31}){\bf e}_x + 
(A_{12}A_{31} + a_j A_{23}){\bf e}_y +
(A_{12}^2 +a_j^2){\bf e}_z}{(A_{12}^2 -|a_j|^2)^2 +
(A_{12}^2+|a_j|^2)(A_{23}^2+ A_{31}^2)]^{1/2}}\; ,
$$
$$
a_{1,2} = \pm i\sqrt{A_{12}^2 + A_{23}^2 + A_{31}^2} \; , 
\qquad a_3 = 0 \; .
$$

The guiding center for the spatial variable ${\bf r}$ is described 
by Eq. (10) in which one has to substitute the form (14) and to average 
the right-hand side according to the rule
$$
\lim_{\tau\rightarrow\infty} \; \frac{1}{\tau} \int_0^\tau \ll 
f({\bf r},{\bf h}(t),{\vec\xi},t)\gg \; dt \; .
$$
In this way, one obtains the equation
\be
\frac{d^2{\bf r}}{dt^2} = \om_1^2 \left ({\bf F} + {\bf G}\right ) \; ,
\ee
in which
$$
{\bf F} = C_3\lim_{\tau\rightarrow\infty} \; \frac{1}{\tau}
\int_0^\tau \left ( b_3^x{\bf e}_x + b_3^y{\bf e}_y +
\lbd b_3^z{\bf e}_z\right ) dt \; ,
$$
$$
C_3 = \frac{(x+h_x^0)S_0^x + (y+h_y^0)S_0^y + \lbd z S_0^z}
{[(x+h_x^0)^2 + (y+h_y^0)^2 +\lbd^2 z^2]^{1/2}}  \; ,
$$
$$
{\bf b}_3 = 
\frac{(x+h_x){\bf e}_x + (y+h_y){\bf e}_y + \lbd z{\bf e}_z}
{[(x+h_x)^2 + (y+h_y)^2 +\lbd^2 z^2]^{1/2}}  \; ,
$$
where $h_\al^0\equiv h_\al(0)$ and $S_0^\al\equiv S^\al(0)$. In the 
integration, defining the effective force ${\bf F}$, one should keep in
mind the occurrence of another small parameter
\be
\left | \frac{\om_1}{\om}\right | \ll 1 \; ,
\ee
because of which ${\bf r}$ is to be treated as a quasi-invariant  with
respect to ${\bf h}$. Taking, for concreteness, the rotating field
\be
h_x(t) = \cos\om t \; , \qquad h_y(t) =\sin\om t 
\ee
results in the effective force
\be
{\bf F} = \frac{[(1+x)S_0^x + yS_0^y + \lbd z S_0^z]
(x{\bf e}_x +y{\bf e}_y +2\lbd^2 z{\bf e}_z)}
{2[(1+2x+x^2+y^2+\lbd^2 z^2)(1+x^2+y^2+\lbd^2 z^2)]^{1/2}} \; .
\ee
We have also considered other forms of the transverse field ${\bf h}$, 
provided it satisfies conditions (12), (13), and (16), which yields 
slightly different formulas for the effective force ${\bf F}$. However, 
slight variations of ${\bf F}$ do not qualitatively change the dynamics 
of atoms. Therefore in what follows, we particularize the effective force 
by Eq. (18).

At this point, it is worth stressing that the force (18), caused by 
magnetic fields, acts on atoms only inside the trap, while outside of 
the latter the force vanishes. Hence, to consider the motion of atoms in 
the whole space, including the state when they fly out of the trap, it is 
necessary to concretize the trap shape. For instance, a cylindrical trap
of radius $R$ and length $L$, measured in units of $R_0$, can be characterized 
by one of the shape factors
\be
\Xi({\bf r}) = 1 -\Theta \left (x^2+y^2 -R^2\right )\;
\Theta\left (|z| -\; \frac{L}{2}\right ) \; , \qquad
\Xi({\bf r}) = \exp\left ( -\; \frac{x^2+y^2}{R^2}\; - \; 
\frac{z^2}{L^2}\right ) \; ,
\ee
in which $\Theta(\cdot)$ is a unit step function. Thus, the general 
expression of the force, acting on atoms inside as well as outside the 
trap, is Eq. (18) multiplied by a shape factor from Eq. (19). It could be 
possible to introduce more refined shape factors. However, to study the 
basic role of the trap finiteness, it is sufficient to consider the simple 
expressions (19). Generally, the shape factors influence atomic motion only 
in the intermediate region, at the instant the atoms leave the trap. In 
this region, the effective force (18) of trapping fields smoothly diminishes
to zero. Because of this, an exact form of a shape factor does not change 
much in the principal behaviour of semiconfined atoms whose main properties
remain practically not altered. In our numerical calculations, we have 
studied the difference resulting from the usage of the step-like and smooth 
shape factors from Eq. (19). For the same trap parameters $R$ and $L$, the 
difference in the atomic trajectories is practically unnoticeable. A slight
distinction can only be noticed in the phase portraits for velocities: The
step-like shape factor causes the appearance of characteristic kinks that are
rounded off by the smooth shape factor. 

To complete the formulation of the problem, we have to fix the initial 
spin polarization ${\bf S}_0\equiv{\bf S}(0)$. As is shown in Refs. [7,8], 
for creating directed motion of atoms, one has to fix the initial spin 
polarization so that
\be
S_0^x = 0 \; , \qquad S_0^y = 0 \; , \qquad S_0^z = S \; .
\ee
Such initial conditions for spin polarization can be prepared in one 
of the ways well known in quantum mechanics [19] or by a short resonant 
pulse, as is discussed in Ref. [20]. Accepting the initial condition (20) 
and taking account of the shape factor (19), we come to the expression 
for the effective force
\be
{\bf F} =\frac{1}{2} \; \lbd S uz \left ( x{\bf e}_x + y{\bf e}_y +
2\lbd^2 z {\bf e}_z \right )
\ee
where $u=u({\bf r})$, with
\be
u({\bf r}) \equiv \frac{\Xi({\bf r})}
{[(1+2x+x^2+y^2+\lbd^2 z^2)(1+x^2+y^2+\lbd^2 z^2)]^{1/2}} \; .
\ee
As is easy to notice, the force (21) is invariant under the transformation
\be
\lbd \rightarrow - \lbd \; , \qquad S\rightarrow - S \; .
\ee
Therefore, it is always possible to choose such $\lbd$ and $S$ that 
$\lbd S>0$, which is assumed in what follows.

It is convenient to pass to the dimensionless time measured in units of
$(\sqrt{\lbd S}\; \om_1)^{-1}$. The return to the dimensional time is 
made by means of the replacement $t\rightarrow\sqrt{\lbd S}\;\om_1 t$. 
Also, let us introduce the dimensionless gravitational force
\be
{\vec \delta}\equiv \frac{{\bf G}}{\lbd S} = 
\frac{{\bf g}}{\lbd S R_0\om_1^2} =\{\delta_\al\} \; .
\ee
Then the evolution equation (15) can be written as the system of equations
$$
\frac{d^2 x}{dt^2} = \frac{1}{2}\; uzx +\delta_x \; , \qquad
\frac{d^2y}{dt^2} =\frac{1}{2}\; uzy +\delta_y \; ,
$$
\be
\frac{d^2 z}{dt^2} = \lbd^2 uz^2 +\delta_z \; ,
\ee
where the function $u=u({\bf r})$ is defined in Eq. (22), with the shape
factor (19).

We accomplished numerical solution of Eqs. (25) for different 
field-anisotropy parameters $\lbd$ and for different trap radii $R$ and
lengths $L$. The results are presented in Figs. 1 to 4, where the $z$ axis 
is along the trap axis and the latter is inclined by the 45 degrees to
the horizon, so that $\delta_z=-\delta_x$. Initial conditions for the
spatial position of atoms are close to the trap center ${\bf r}_0=0$,
and initial velocities of atoms, $\dot{\bf r}_0$, are varied in the 
interval $-0.1\leq\dot{r}_0^\al\leq 0.1$. The motion $x(t)$ is similar
to $y(t)$, because of which we show only the $x-z$ cross-section of phase
portraits. The pictures do not qualitatively change when varying the trap
parameters $R$ and $L$ in the interval $[1,10]$. But the acceleration 
is better for longer traps, with larger $L$, as well as for larger $\lbd$. 
The maximal velocity that can be achieved is of the order of 
$v_{max}\sim \lbd\sqrt{L}$. The aspect ratio $R_{asp}=<x>/<z>$ is small,
$R_{asp}\sim 0.01$, showing that the beam is stretched in the $z$-direction 
about $100$ times larger than in the $x$-direction, that is, the beam is
well collimated. The figures present calculations for the step-like shape 
factor which, to our mind, more clearly illustrates the role of the trap 
finiteness. As we have numerically checked, the usage of the smooth factor,
with the same trap parameters, does not alter atomic trajectories. The sole 
difference appears in the phase portraits showing the axial to radial 
velocities: In the case of the step-like shape factor there are kinks in 
the intermediate region where atoms leave the trap. These kinks are rounded 
off in the case of the smooth shape factor.

Comparing these numerical calculations with those of Ref. [8], where gravity 
was not taken into account, we come to the following conclusions. The regime
of motion, called {\it semiconfinement}, remains, when atoms are confined 
from one side of the trap by the minimal value
$$
z_{min} \approx -\left ( \frac{6}{\lbd^2} \; \dot{z}_0^2\right )^{1/3} \; ,
$$
but are not confined from another side, escaping with acceleration 
in this direction. For the present choice of coordinates, the escape 
direction is along the axis $z$. If the gravity component $\delta_z>0$, 
the positive gravitational force only helps to atoms to move preferably 
in the $z$-direction, so that all atoms are semiconfined and escape from 
the trap forming a narrow beam along the $z$-axis. However, if 
$\delta_z<0$, then gravity hampers to escape for some of the atoms. 
The fraction of atoms that remain confined because of gravity is of order 
$\frac{1}{2}\lbd\sqrt{\frac{1}{3}|\delta_z|}$. The main visible influence 
of gravity on the ejected beam is in curving atomic trajectories in the 
same way as gravity curves the trajectory of a ball shot out of a cannon.

To have a feeling of the values of the characteristic quantities, met 
above, let us give estimates for the case of parameters typical of alkali 
atoms in magnetic traps [1--3,9,21]. The characteristic frequency of atomic 
motion is $\om_1\sim 10^2-10^3$ s$^{-1}$, that of spin motion is 
$\om_2\sim 10^7-10^8$ s$^{-1}$. The frequency of the transverse rotating 
field is $\om\sim 10^4-10^5$ s$^{-1}$. The collision rate is $\gm\sim 10$
s$^{-1}$. Consequently,
$$
\frac{\om_1}{\om_2} \sim 10 ^{-5} \; , \qquad
\frac{\om}{\om_2} \sim 10 ^{-3} \; , \qquad
\frac{\gm}{\om_2} \sim 10 ^{-6} \; , \qquad
\frac{\om_1}{\om} \sim 10 ^{-2} \; , 
$$
from where it follows that the inequalities (12) and (16) hold true. The
characteristic radius (6) is $R_0\sim 0.1 -0.5$ cm. The dimensionless
gravity components, defined in Eq. (24), can always be made small. Thus,
for $\lbd \sim 2$, $S\sim 1$, $R_0\sim 0.5$ cm, and $\om_1\sim 10^2-10^3$
s$^{-1}$, substituting in Eq. (24) the gravitational acceleration 
$g^\al\sim 10^3$ cm/s$^2$, we get $\delta_\al\sim 10^{-3} - 10^{-1}$.
When the trap axis is directed against the gravitational force, then
the amount of atoms that cannot escape from the trap is between $1\%$
to $10\%$. Hence, the semiconfining regime  of motion is not much 
spoiled by gravity. The majority of atoms fly out of the trap, forming 
a well-collimated beam with a squeezing factor of about 100; the maximal 
velocity of atoms, for a trap of length $1-10$ cm can reach the order 
of 100 cm/s.

The role of random pair collisions can be considered in the same way as 
in Ref. [8]. These collisions do not essentially disturb the semiconfining
regime provided that
$$
\frac{\gm^2 D}{(\lbd S)^{3/2}\om_1^3} \ll 1 \; , \qquad
D\equiv \sup_\al\{ D_\al\} \; .
$$
This condition, accepting for the diffusion rate an estimate 
$D\sim k_B T/\hbar$, where $k_B$ is the Boltzmann constant and $T$ is
temperature, can be presented as
$$
T \ll T_c \equiv \frac{\hbar(\lbd S)^{3/2}\om_1^3}{k_B\gm^2} \; .
$$
These inequalities show that the disorganizing influence of atomic 
collisions is suppressed when increasing the anisotropy parameter $\lbd$
or lowering temperature. For the characteristic values $\lbd S\sim 1$,
$\om_1\sim 10^2-10^3$ s$^{-1}$, and $\gm\sim 10$ s$^{-1}$, the critical
temperature, below which the semiconfining regime appears, is 
$T_c\sim 10^{-7}-10^{-4}$ K. This implies that for successful functioning
of an atom cannon, trapped atoms are to be sufficiently cooled down. Such 
a cooling can be effectively realized in modern traps [1--3].

\section{Discussion}

In this paper, we have analysed the influence on the magnetic
semiconfinement of neutral atoms of three factors, trap size, field 
anisotropy, and gravity, which had not been considered in the previous
works [7,8]. The principal result is that, despite the action of gravity, 
the semiconfining regime can always be preserved by appropriately choosing 
trap parameters. Consequently, the suggested magnetic semiconfinement can 
really serve as an efficient mechanism for creating well-collimated beams
of neutral particles and atoms.

In order to correctly describe the semiconfining regime of motion, it has 
been necessary to lift the adiabatic approximation that is usually employed for
considering the motion of trapped atoms. As is known from the general 
theory [17], the adiabatic approximation may be reasonable for treating the 
motion of a multifrequency dynamical system that is close to a stationary 
state. The motion of atoms, permanently confined inside a trap, is exactly 
such a kind of a stationary regime. This is why the adiabatic approximation 
works reasonably well for constantly confined atoms. Corrections to this 
approximation, in the case of a TOP trap, can be estimated as being of the 
order of $\om_1/\om\sim 10^{-2}$, which being expressed in percent, makes at 
most a few percent. Atomic micromotion of atoms in a TOP trap has recently 
been studied experimentally [22,23]. The detected nonadiabatic effects were 
found to be of the order of a few percent, in agreement with the above 
estimate. Anharmonic corrections to the harmonic approximation of the 
effective TOP potential are also not large for trapped atoms moving close 
to the trap center [24].

The situation is different for atoms accelerated out of trap in the 
semiconfining regime of motion. Such a regime is far from stationary, 
because of which the adiabatic approximation is not applicable in principle 
[17]. Also, in this regime, atoms may move far from the trap center, which 
prohibits the usage of a harmonic approximation. This is why, we have 
considered the effective trap force (21) in its complete form, containing 
the factor (22), never invoking harmonic approximations. At the same time, 
taking into account the anharmonic factor (22) not only makes the consideration 
more general but also reduces the influence of a particular shape factor on 
the motion of atoms. This happens because the role of the shape factor becomes 
noticeable when atoms leave the trap. But the effective force (21), due to
the anharmonic factor (22), fastly diminishes to zero after atoms leave the 
death circle whose radius is usually essentially smaller than the trap size.

As is explained in the last paragraph of the previous section, the 
semiconfining regime is easier realized after precooling to sufficiently 
low temperatures. In the region of the estimated temperatures, 
Bose-Einstein condensates can be created. Atomic beams outcoupled from 
condensates remind photon beams from lasers, because of which a device 
emitting condensed atoms can be called an atom laser. The known standard 
way for realizing an output coupler for trapped atoms is by transferring 
atoms from a trapped state into an untrapped state by means of a 
monochromatic resonant rf field, as has been done in experiments [25--27]. 
This method produces a beam of atoms merely falling down in the field of 
gravity. Such a device can hardly be called an atom laser, since the very 
first condition on a laser is that its output could be pointed in an 
arbitrary direction. A directional extraction of sodium atoms from a 
trapped condensate was demonstrated by using two laser beams stimulating 
Raman transitions [28]. More information on atom lasers can be found in 
reviews [2,3]. The mechanism for creating atomic beams, we suggest, differs 
from the mentioned above in three main points: (i) it can be realized by 
using only magnetic fields, without involving additional blocking devices 
or external laser beams; (ii) the beam direction is prescribed by the trap 
axis that can be oriented arbitrarily; (iii) the characteristics of the 
semiconfinement can be varied by choosing appropriate magnetic fields. 
In this way, the proposed mechanism can be employed for creating 
highly-directional beams from atom lasers.

It would, certainly, be nice if the effect of semiconfinement could be
observed experimentally. Since we are not experimentalists, it would not be 
reasonable from our side to plunge into a discussion of technical details 
of experiments. Our aim has been to demonstrate the principal feasibility 
of realizing the semiconfining regime of motion. However, to motivate 
experimentalists, it would, probably, be useful to touch several points 
whose clarification could facilitate experimental work.

First of all, we would like to emphasize that the described effect is rather 
general and can be realized in different traps, with different magnetic fields. 
A dynamic TOP-like trap was considered above for concreteness. As we have 
checked, the semiconfining regime of motion exists for static traps as well, 
provided the required initial conditions for atomic spins are prepared. We 
could imagine at least three ways to achieve the necessary initial spin 
polarization.

One possibility could be to orient spins in the $z$-direction by means of a 
guiding longitudinal field, while the transverse field is yet switched off.
After this, the latter field should be quickly switched on. The main 
problem here is that the trap field has to be switched on faster than the 
Larmor period $2\pi/\om_2$. This looks to be difficult to accomplish 
experimentally. Although, what seems difficult today may be successfully 
performed in future.

Another possibility is to prepare, first, atoms with the required 
$z$-polarization of spins in a trap. For instance, this can be done by 
using the trap of Ref. [29], which is a quadrupole trap with a bias field 
along the $z$-axis. Then the spin-polarized atoms are to be quickly loaded 
into another trap with the desired field configuration. The feasibility of 
transferring atoms from one trap to another by means of a sudden transfer, 
as opposed to slow transfer, has been discussed in Ref. [30].

One more way could be as follows. Assume that a trap consists of two 
chambers, upper and lower. Let atoms be confined in the upper chamber. 
Then, acting on these atoms by rf field, one transfers them into an 
untrapped state [25--28]. Obtaining the spin polarization 
corresponding to this untrapped state, atoms fall down because of gravity, 
and pass to the lower chamber. The field configuration of the latter is to 
be such that the spin polarization of the falling atoms be that required 
for the semiconfining regime of motion. Thus, the lower chamber could emit 
a quasi-continuous beam of atoms.

The features of the emitted atomic beam depend, of course, on many 
technical details related to particular experiments. The major properties 
are prescribed by the magnetic field configuration arranged. The trap shape 
factor, as is explained above, plays less important role, when the trap 
size is larger than the radius of the death circle or the radius of an 
atomic cloud in a trap. A continuous atomic beam being ejected out of the 
trap will be influenced by the trap shape slightly stronger than a short 
pulse. In the latter case, when the group of atoms, after an initial 
acceleration reaches the trap boundary, one could switch off the trap field 
abruptly. Such a situation would exactly correspond to the usage of the 
step-like shape factor.

Finally, one could imagine experiments for which the directed motion of 
atoms, organized in the proposed way, would not necessarily end up with 
atoms leaving the trap. Some of such possibilities are mentioned in the 
Introduction. Probably, the most interesting suggestion would be to study 
the relative motion of one atomic component through another, as is 
discussed in Ref. [7].

\vskip 5mm

{\bf Acknowledgement}

\vskip 2mm

We are grateful for useful remarks to V.S. Bagnato and A.B. Kuklov.
We acknowledge financial support by the Bogolubov-Infeld Grant of the 
State Agency for Atomic Energy, Poland.

\newpage

\begin{center}

{\bf Figure Captions}

\end{center}

\vskip 2cm

{\bf Fig. 1}. The trajectories of atoms in the $x-z$ cross-section 
at the initial stage of acceleration, when $0\leq t\leq 50$, for the 
trap parameters $R=1,\; L=1,\; \lbd=20$, and the gravity components
$\delta_x=0.01,\; \delta_y=0,\;\delta_z=-0.01$.

\vskip 5mm

{\bf Fig. 2}. The velocities of atoms on the $\dot{x}-\dot{z}$ plane for
the same conditions as in Fig. 1. The characteristic kinks is a result of a 
step-like shape.

\vskip 5mm

{\bf Fig. 3}. Typical trajectories during the long time interval 
$0\leq t\leq 300$ for the trap parameters $R=1,\; L=1,\; \lbd=10$, and 
the gravity components $\delta_x=0.05,\; \delta_y=0,\; \delta_z=-0.05$.

\vskip 5mm

{\bf Fig. 4}. The phase portrait for atomic velocities corresponding to 
the parameters of Fig. 3. The kinks are again due to a step-like shape 
factor. The general picture for a smooth shape factor remains the same, 
except that the kinks become rounded off.

\end{document}